\documentstyle[11pt]{article}
\input epsf

\newlength{\awidth}
\newlength{\aheight}
\newlength{\uswidth}
\newlength{\usheight}
\def\preprint#1{\gdef\@preprint{#1}}
\def\bce{\begin{center}}
\def\ece{\end{center}}
\def\be{\begin{equation}}
\def\ee{\end{equation}}
\def\bea{\begin{eqnarray}}
\def\eea{\end{eqnarray}}

\newcounter{fignr}
\newenvironment{fig}[1]{\refstepcounter{fignr}\label{#1}\begin{center}}{
    \end{center}}
\newcommand{\figcap}[2]{\parbox{#1}{{\footnotesize  Fig. \thefignr. #2}}}

\begin{document}
\baselineskip=.285in

\catcode`\@=11
\def\maketitle{\par
 \begingroup
 \def\thefootnote{\fnsymbol{footnote}}
 \def\@makefnmark{\hbox
 to 0pt{$^{\@thefnmark}$\hss}}
 \if@twocolumn
 \twocolumn[\@maketitle]
 \else \newpage
 \global\@topnum\z@ \@maketitle \fi\thispagestyle{empty}\@thanks
 \endgroup
 \setcounter{footnote}{0}
 \let\maketitle\relax
 \let\@maketitle\relax
 \gdef\@thanks{}\gdef\@author{}\gdef\@title{}\let\thanks\relax}
\def\@maketitle{\newpage
 \null
 \hbox to\textwidth{\hfil\hbox{\begin{tabular}{r}\@preprint\end{tabular}}}
 \vskip 2em \begin{center}
 {\Large\bf \@title \par} \vskip 1.5em {\normalsize \lineskip .5em
\begin{tabular}[t]{c}\@author
 \end{tabular}\par}
 \end{center}
 \par
 \vskip 1.5em}
\def\preprint#1{\gdef\@preprint{#1}}
\def\abstract{\if@twocolumn
\section*{Abstract}
\else \normalsize
\begin{center}
{\large\bf Abstract\vspace{-.5em}\vspace{0pt}}
\end{center}
\quotation
\fi}
\def\endabstract{\if@twocolumn\else\endquotation\fi}
\catcode`\@=12

\preprint{}
\title{\Large\bf Anomalous Center of Mass Shift: \\ Gravitational Dipole Moment
\protect\\[1mm]\  }
\author{\normalsize Eue Jin Jeong\\[1mm]
{\normalsize\it Department of Physics, Natural Science Research Institute, 
Yonsei University, Seoul, Korea}}

\maketitle

\renewcommand{\theequation}{\thesection.\arabic{equation}}
\def\gatij{\gamma^{ij}}
\def\gabij{\gamma_{ij}}
\def\ophi{\phi^{a}}
\def\hphi{\overline{\phi}^{a}}

\begin{center}
{\large\bf Abstract}\\[3mm]
\end{center}
\indent\indent
The anomalous, energy dependent shift of the center of mass of an 
idealized, perfectly rigid, uniformly rotating hemispherical shell 
which is caused by the relativistic mass increase effect is 
investigated in detail. It is shown that a classical object on 
impact which has the harmonic binding force between the adjacent 
constituent particles has the similar effect of the energy dependent, 
anomalous shift of the center of mass. From these observations, 
the general mode of the linear acceleration is suggested to be 
caused by the anomalous center of mass shift whether it's due to 
classical or relativistic origin. The effect of the energy dependent 
center of mass shift perpendicular to the plane of rotation of a 
rotating hemisphere appears as the non zero gravitational dipole 
moment in general relativity. Controlled experiment for the 
measurement of the gravitational dipole field and its possible 
links to the cylindrical type line formation of a worm hole in the 
extreme case are suggested. The jets from the black hole accretion 
disc and the observed anomalous red shift from far away galaxies 
are considered to be the consequences of the two different aspects 
of the dipole gravity.  
\vspace{1cm}\noindent

Keyword(s): gravitational dipole moment, center of mass shift, general
relativity, worm hole, anomalous red shift, jets from black hole 
accretion disc.

\noindent
PACS number(s): 03.30.+p, 04.30.+x

\newpage
\baselineskip=15pt

\pagenumbering{arabic}
\thispagestyle{plain}
\setcounter{section}{1}
\begin{center}\section*{\large\bf I. Introduction}\end{center}
\indent\indent
One of the puzzling issues in gravitational theory is how one can 
generate a directional gravitational field to produce linear 
acceleration. Conjectures and speculations abound on the topics 
of space travel, worm hole and time machines\cite{B1}\cite{B2}
\cite{B3}\cite{B4}. However, the problems largely remain unsolved 
and the key mechanisms have been awaiting to be unfolded. In this 
paper, the anomalous shift of the center of mass of an idealized 
perfectly rigid spinning axisymmetric object which has the point 
asymmetry with respect to the center of mass is investigated in 
detail, which may help shed a light on this particular issue. 

\setcounter{section}{2}
\bce\section*{\large\bf II. Anomalous Center of Mass Shift}
\ece
\indent
\bce\subsection*{1. Hemispherical Rotor}
\ece

Consider an infinitesimal moment of time $\delta$t during which 
a measurement is made on the mass of the mass components $m_{i}$ 
forming an idealized perfectly rigid hemispherical shell 
(Fig.~\ref{fig1}) placed in an asymptotically flat spacetime 
region at rest except one angular rotational degree of freedom 
along the symmetry axis. The system is located in the empty 
space totally at rest initially and then a slight tap is applied 
at the rim of the rotor in the tangential direction perpendicular 
to the rotational symmetry axis to impart a torque to the system. 
The system has reached a uniform angular frequency of rotation 
$\omega$. The mass of the component $m_{i}$ is observed to be 
increased by  \\ \\
\[m_{i}^{*} = 
\frac{m_{i}}{\sqrt {1 - \frac{\omega^{2}r_{i}^{2}}{c^{2}}}}\]\\ \\
by the special relativistic effect where $r_{i}$ is the distance 
from the rotation axis to the position where $m_{i}$ is located, 
and $\omega$ the angular frequency of the rotor, where the 
hemispherical rotor is assumed to be made of such an ideal 
material that it retains its original shape even at extremely 
high rotational velocities so that $r_{i}$ remains fixed 
independent of $\omega$. The center of mass is defined as a 
point in the space where the total mass of a system of objects 
is regarded to be concentrated by the rest of the universe at 
the instant moment of measurement. The center of mass can be 
expressed in terms of a summation of the contributions from the 
individual $m_{i}$s as follows (see Fig.~\ref{fig1}):\\ \\
\[\overline{r}_{c} = \frac {\sum {m_{i}^{*}
\overline{r}_{i}}}{\sum{m_{i}^{*}}}\]	\\ \\
If the mass elements $m_{i}$  and therefore $m_{i}^{*}$ are made 
sufficiently small, this expression can be approximated by the 
integral form \\ \\ 
\[\overline{r}_{c} = \frac {\int \overline{r}\delta m^{*}}{\int \delta
m^{*}}\] \\ \\
in the limit the elements of mass $m_{i}$ approaches zero. This method is 
identical to that of the Thirring's work\cite{B5} on the rotating spherical 
mass shell where the integral has been performed in the rest frame 
of the rotor by employing the four velocity, length contraction and 
the constant mass density $\sigma$. 

For the axisymmetric hemispherical case shown in Fig.~\ref{fig1}, 
only the z component of the center of mass is non zero, and it is 
explicitly given by \\ \\
\[r_{c} = \frac{2 \pi \sigma R^{2} 
 \int_{-\frac{\pi}{2}}^{0} \frac{-R \cos {\theta} \sin{\theta} \,d\theta}
 {\sqrt{1 - \frac {\omega^2 R^2 \sin^2{\theta}}{c^2}}}}
 {2 \pi \sigma R^{2} 
 \int_{-\frac{\pi}{2}}^{0} \frac{- \sin{\theta} \,d\theta}
 {\sqrt{1 - \frac {\omega^2 R^2 \sin^2{\theta}}{c^2}}}} = 
 \frac {- R \frac {1 - \sqrt{1 - \alpha}}{\alpha}}
 {\sqrt{\frac {1}{\alpha}} \sinh^{-1}{\sqrt{\frac{\alpha}
{1 - \alpha}}}}\] \\ \\
where \[\alpha = \frac {\omega^2 R^2}{c^2}\]\\ \\
and $\sigma$ is the inertial mass per unit area of the 
hemispherical shell with uniform thickness. Depending on the 
angular speed and consequently the rotational kinetic energy 
of the rotor, the system can have its center of mass at any 
position between  R/2 and 0. It is a smooth function of 
$\alpha$ comprising the center path of the curve shown in 
Fig.~\ref{fig2}. Note that the usual rotational kinetic 
energy of a spinning object has become like the ``potential'' 
energy since the energy has developed a dependency on the 
physical length. The ``anomalous center of mass shift'' is 
defined as the distance between the center of mass of an 
energetically excited system at a given moment of time and 
that of the assumed ground energy state of the same system 
at the corresponding moment of time. 
\ \\
\noindent\parbox{\textwidth}{\noindent\begin{fig}{fig1}
  \mbox{\setlength{\epsfxsize}{.80\textwidth} \epsfbox{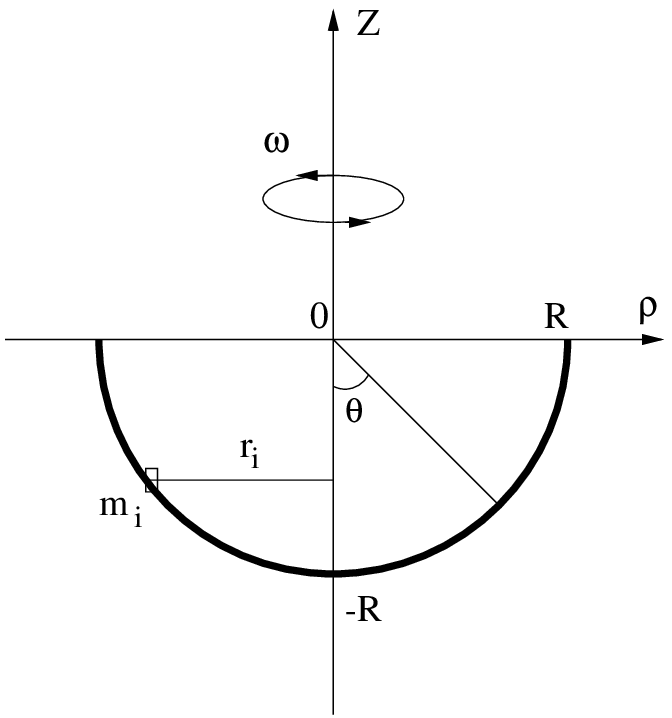} }\\
\ \\
  \figcap{.9\textwidth}{A hemispherical shell of radius R rotating 
with the angular speed $\omega$. Mass component $m_{i}$ is located 
at $r_{i}$ from the rotation axis.}
\end{fig}
}\\

Therefore, the anomalous center of mass shift $\delta r_{c}$
for the hemispherical shell shown in Fig.~\ref{fig1}. is given by \\ \\
\[\delta r_{c} = r_{c} - r_{o} = \frac{R}{2} -
\frac {- R \frac {1 - \sqrt{1 - \alpha}}{\alpha}}{\sqrt{\frac {1}{\alpha}}
\sinh^{-1}{\sqrt{\frac{\alpha}{1 - \alpha}}}}\] \\ \\
where $r_{o}$ is the ground state center of mass with zero angular
 speed of rotation. It is a very slowly increasing function of 
$\alpha$ which is approximately given by \\ \\ 
\[\delta r_{c} = \frac{\omega^2 R^3}{24c^2}\]\\ \\
for $R\omega << c$. The above length element defined as the 
``anomalous center of mass shift'' arising from the energy dependent 
center of mass has the following characteristics. 
1. It is related to an internal energy of the system. 
2. The larger the shift of the center of mass, the greater the
stored energy. 
3. It can be returned to zero upon releasing the related internal energy.

According to Newtonian mechanics\cite{B6},
shifting the center of mass of an object without impressed external 
action in the direction of the movement of the object is a 
nonsensical  proposition. However, since the relativistic mass 
increase effect has been experimentally proven to be correct to 
a high degree of accuracy, the anomalous shift of the center of 
mass presented above must be a physically observable effect. This 
apparently represents a case of violating Newton's first and third 
law of motion since the spinning hemisphere is capable of 
spontaneously developing a displacement of its own center of 
mass perpendicular to  the plane of rotation independent of the 
choice of the coordinate system. Even if we take into account 
the external source of the force that has given the torque to 
the system, the direction of the center of mass shift is not 
consistent with the force applied for the torque since they are 
perpendicular to each other.  Consequently, we face a conflict 
between Newtonian mechanics and special relativity apart from the 
well documented problem of the frame of reference. 

Here, we are forced to choose the path of the notion that special 
relativity is correct and that there are cases in which an object 
experiences the shift of its center of mass without impressed action, 
in stark contradiction to the Newton's first and third law of
motion. Once we follow such path, Newton's first and third law of 
motion must be regarded as a special case of the generalized version 
which will be stated shortly since this system represents an 
unequivocal exception to these laws in the form in which Newton 
originally cast them .

\ \\
\noindent\parbox{\textwidth}{\noindent\begin{fig}{fig2}
  \mbox{\setlength{\epsfxsize}{.80\textwidth} \epsfbox{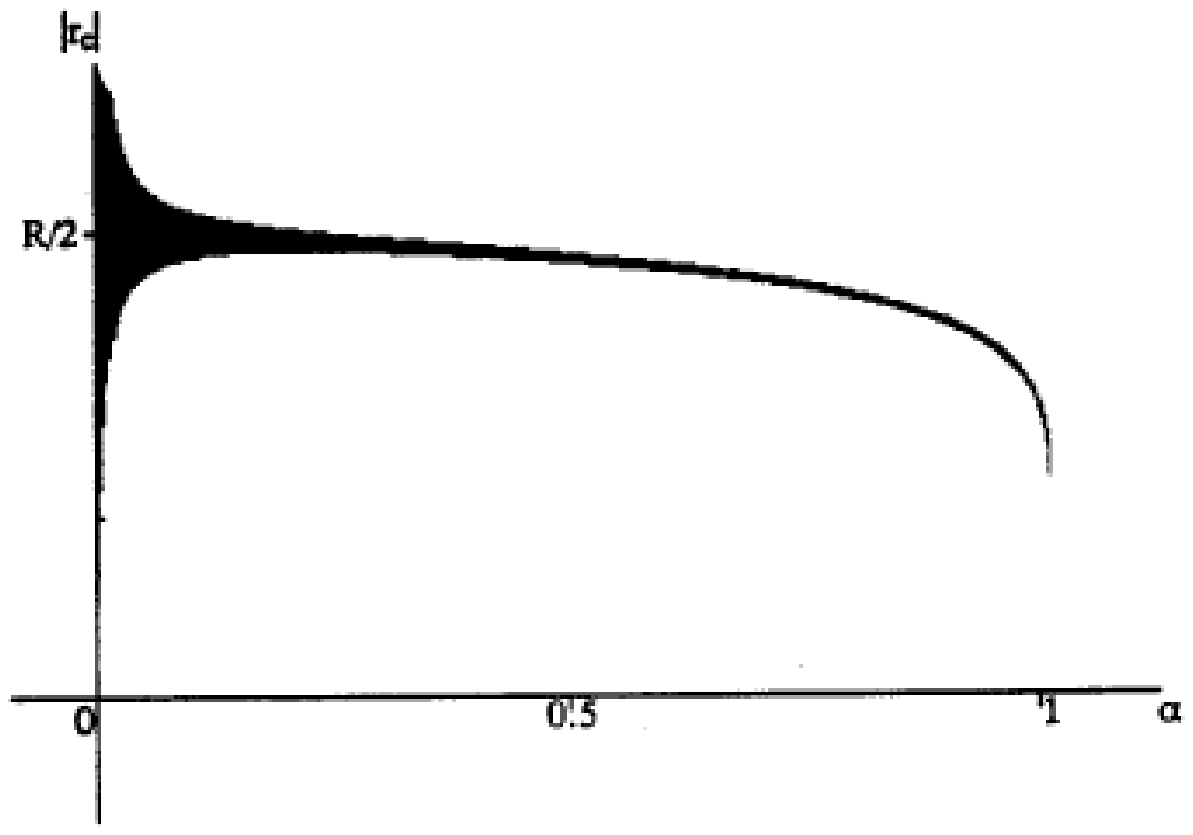} }\\
\ \\
  \figcap{.9\textwidth}{The anomalous relativistic center of mass 
of a spinning hemispherical shell (Fig.~\ref{fig1}) as a function 
of $\alpha$ (computer generated based on the equation for $r_{c}$). }
\end{fig}
}\\

It is noted that neither Newtonian mechanics nor the presently 
known results of general relativity has the scope of handling this 
phenomenon. It is clear that this idealized hemispherical system 
exhibits a peculiar mechanical property which requires close
scrutiny. It is also noted that this property is totally due to the 
specific geometrical configuration of the hemisphere which has 
the longitudinal axially asymmetric shape in which the individual 
mass components are constrained to rotate collectively. Although the 
individual mass components do not exhibit anomalous physical 
effect in isolation, the whole system does. We have seen many 
times in physics that whenever the symmetry of a physical system 
breaks down, there appear new physical phenomena. It can also be
seen easily that the translational gauge symmetry in general 
relativity is broken in this system. The usual constant phase 
factor that used to shift the coordinate system without affecting 
the system's energy content can not be a constant anymore in
the present case. It has to carry the information about the 
rotational kinetic energy of the source which depends on the 
time derivative of the angular orientation ($w = \,d\theta/\,dt$) 
of the rotor to know the exact effective center of mass of the 
system. This breaking of translational gauge symmetry also 
implies the breaking of energy and momentum conservation law 
which are the direct consequences of the Newton's laws of
motion. The crucial question here may be ``How would the system 
behave by having created its anomalous center of mass shift ?''

\bce\subsection*{2. Mechanics of a Ball on Impact}
\ece

To investigate the physical significance of this shift further, 
consider a classical example of an elastic metallic ball at 
rest hit by a bat instantaneously during the infinitesimal moment 
of time $\delta t$. The location of each atomic component of
the mass $m_{i}$ is shifted by - $\delta \overline{r}_{i}$ at 
the moment of impact. The over all shift of the center of mass 
of the ball is expected to be in the opposite direction to that 
of the impact since the inertia resists to change its position. 
The reason for this is also because 
the bat contributes temporarily to the total mass of the body 
at the moment the impact is applied. Therefore, The shifted center of mass 
of the ball at 
time $t = 0$  is given by \\ \\
\[\overline{r}_{c} = \frac {\sum{m_{i} (\overline{r}_{i} - \delta
\overline{r}_{i})}}{\sum{m_{i}}}\] \\ \\
while \\ \\
\[\overline{r}_{o} = \frac {\sum{m_{i} \overline{r}_{i}}}{\sum{m_{i}}}\] \\ \\
and \\ \\
\[\delta \overline{r}_{c} = 
\frac {\sum{m_{i} (\overline{r}_{i} - \delta \overline{r}_{i})}}{\sum{m_{i}}} 
- \frac {\sum{m_{i} \overline{r}_{i}}}{\sum{m_{i}}}
= \frac { - \sum{m_{i} \delta \overline{r}_{i}}}{\sum{m_{i}}}\] \\ \\

The $\delta r_{c}$ is the momentary internal shift of the center 
of mass at the moment the ball is subjected to the impact. The 
motion of the ball at the later time $t_{o}$ is obvious 
from our day to day experiences. The internal center of mass 
recovers its original position instantly and the ball moves at 
a constant speed in the direction in which the temporary shift 
of the center of mass has occurred. 

In classical mechanics, the impact \\ \\
\[\overline{F}\Delta t = \Delta \overline{P}\] \\ \\
is given to the ball by the step function\\ \\ 
\[\begin{array}{lcr}
\begin{array}{lll}
 \Delta \overline{P}(t) & = & 0 \\
 & = & \overline{F}\Delta t \\
\end{array} &
\mbox{\hspace{1cm}}&
\begin{array}{l}
 t < 0 \\
 t \geq 0 \\
\end{array}
\end{array}\] \\ \\
and the force is given by the delta function \\ \\
\[\overline{F}(t) = \frac{\Delta \overline{P}}{\Delta t} \delta(t) =
\overline{F}\delta(t)\] \\ \\
Now, the force $\overline{F}$ must be related to the sum of 
the restoring forces of the harmonic 
oscillators in the solid upon impact which is given by \\ \\
\[\overline{F}_{restore} = \sum{k_{i}\delta \overline{r}_{i}}\] \\ \\
The net component of the restoring force toward the direction 
of the impact is written by \\ \\
\[\overline{F}_{restore} = \sum{k_{i}|\delta \overline{r}_{i}|
\cos{\theta_{i}}\hat{x}}\] \\ \\
where $\theta_{i}$ is the angle between $\hat{x}$ and the displacement 
$\delta \overline{r}_{i}$ and $\hat{x}$ the unit vector defined by $\delta
\overline{r}_{c}/ | \delta \overline{r}_{c}|$. 
We can assume this restoring force is the same as the force
$\overline{F}$ acting on the object \\ \\
\[\overline{F} = \overline{F}_{restore} = \sum{k_{i}|\delta \overline{r}_{i}|
\cos{\theta_{i}}\hat{x}}\] \\ \\
with the identification that the $\Delta t$ is the mean relaxation 
time required for the damped harmonic oscillators to return to at 
rest. By scaling down the mass component $m_{i}$ to the 
individual atoms of the solid, it is represented by the atomic 
mass $(m_{i} = m)$ and the force constant $k_{i}$ by the atomic 
force constant. In the limit, the center of mass shift becomes \\ \\
\[\delta \overline{r}_{c} = - \frac {\sum{m_{i} \delta
\overline{r}_{i}}}{\sum{m_{i}}} = - \sum{\delta \overline{r}_{i}}
= - \sum{|\delta \overline{r}_{i}| \cos{\theta_{i}}\hat{x}}\] \\ \\
since the components perpendicular to the direction of the center 
of mass shift are canceled by themselves.

Assuming also that the atomic force constant $k_{i}$ is the same 
for all atomic oscillators $(k_{i} = k)$, the force $\overline{F}$ 
can be written \\ \\
\[\overline{F} = \overline{F}_{restore} = 
\sum{k_{i}|\delta \overline{r}_{i}|
\cos{\theta_{i}}\hat{x}}
= k \sum{|\delta \overline{r}_{i}| \cos{\theta_{i}}\hat{x}}\] \\ \\
Therefore, the acting force is given in terms of the center of 
mass shift by \\ \\
\[\overline{F} = - k\delta \overline{r}_{c}\] \\ \\
and the total impact by \\ \\
\[\overline{P} = \int \overline{F}\,dt = - \int k \delta\overline{r}_{c}(t)\,dt\]
\\ \\
where $\delta \overline{r}_{c}(t)$ is the time dependent center 
of mass shift obtained by solving 
the damped harmonic oscillator with the appropriate initial conditions. 
It is noted that this center of mass shift conforms to the 
definition of the anomalous center of mass shift and also has 
all the subsequent characteristics.

\bce\subsection*{3. Mach's Principle}
\ece

From the above discussions, the following general rules can be 
stated on the dynamics of an object subject to an acceleration. \\
\begin {enumerate}
\item The strength of the force acting on an object is 
proportional to the anomalous center of mass shift of the object.  
\item The acting force is the same as the restoring force 
arising from the anomalous center of mass shift
\end{enumerate}
Note that the motion of an object under an external force is 
described in terms of the anomalous center of mass shift of the 
object perceived by the rest of the universe. 
This statement is in fact Mach's principle written in terms of 
the physically measurable quantities. It is considered that 
these rules are an extension of Newton's first and 
third law of motion since they don't mandate the presence of 
an obvious external force acting on the body only if there 
exists the anomalous center of mass shift perceived by 
the rest of the universe for a reaction to be initiated on 
the object. In 1893 Ernst Mach stated the hypothesis: The 
influence of all the mass in the universe determines what 
is natural motion and how hard it is to change, which was 
labeled ``Mach's principle'' by Einstein. The rest of the 
universe somehow abhors a local system of objects to develop 
the anomalous center of mass shift by forcing it to return to
zero as quickly as possible. 

Newtonian mechanics is based on the assumption that any 
spatially extended object can be conceptually reduced into 
a point mass for the purpose of describing the trajectory 
of the object, where the position of the point mass is 
tacitly assumed to be at the unequivocal, rotation independent 
center of mass of the object. This works well for 
all spherically symmetric objects and any object which has 
axisymmetry and also point symmetry with respect to the 
center of mass, assuming the object is in rotational
motion along the symmetry axis, and also for  non rotating 
systems.  As shown so far, this assumption doesn't work for 
the rotating hemisphere since
the center of mass is not rotation independent. Once the 
hemisphere is reduced into a point mass, it would no 
longer be able to develop the anomalous center of mass shift. 
The reducing process destroys the important mechanical 
characteristic of the object. In fact, the assumption 
would not work in general for objects of arbitrary shape
in arbitrary rotational motion. For an idealized point 
mass for which the rotational degree of freedom is not defined, 
imparting the momentum(either to or from) is
also seen to be conceptually impossible. Such an idealized 
point mass is out of context 
for both the Newtonian system and the Machian presented 
in this paper. Since any spatially extended object must 
have the structure due to the binding force, the 
anomalous center of mass shift becomes a general feature 
of such an object on impact or transfer of momentum in general.

\bce\section*{\large\bf lll. Universal Linear Force}
\ece
\indent\indent
	 
The physical similarity between the two cases lies in the 
fact that the shifted center 
of mass represents an energetically excited state and once 
the stored energy is released the anomalous center of mass 
shift returns to zero for both systems. 
According to the above rules, the hemispherical rotor must 
experience the force toward the shifted center of mass 
proportional to amount of the shift in the center 
of mass, assuming that the laws derived from the motion of 
a ball on impact can equally be applied to the anomalous 
center of mass shift caused by the special relativistic mass 
increase effect.   

To find out if this is indeed the case, one can see the 
picture in a quantitative basis by the following classical 
arguments. The center of mass of the hemispherical rotor at 
the rotational ground energy state $(\omega = 0)$ is 
denoted by RCM and the excited state center 
of mass by NCM as shown in Fig.~\ref{fig3}. 

In classical rotations, the centrifugal force of the rotating 
object is perpendicular to the rotational symmetry axis and 
outward with respect to RCM which is also the center 
of the centripetal force and the total sum of these forces 
acting on the rotor is zero as 
long as the structure remains in equilibrium. The centripetal 
force is exerted by the atomic or molecular binding force of 
solids, on the other hand, the centrifugal force caused by 
inertia is believed to be exerted by the rest of the matter 
in the universe according to Mach. It is noted that these 
two forces have completely different origin from each other. 
And there is no reason to expect that these two forces have to
always cancel each other out in every physical situations. 
For example, if the centripetal force were happen to be 
weaker than the centrifugal force from the rotational motion,
the rotor would be torn apart. 

Tapping the hemispherical rotor at the rim in the tangential 
direction perpendicular to the rotation axis has resulted 
in the longitudinal axial displacement of the center of 
mass of the rotor due to the relativistic mass increase effect. 
Now the question is how an object can effectively move from 
point A to B and stop without any linear force being 
involved in the system. The inertial principle of motion 
suggests it is not possible unless 
some form of force is invoked in the process. 
To find out if there is any force that can be 
the cause of the unknown force, consider the following. 
As the rotation reaches certain uniform angular frequency 
$\omega$ (Fig.~\ref{fig3}), there arises the problem of the 
mismatch between the centripetal and the centrifugal 
force because of the development of the two slightly 
different center of mass, which does not happen in the case 
of a rotating sphere.  While the center of the centripetal 
force is still considered to be at the stationary point RCM, 
the centrifugal force exerts its outward normal force with 
respect to the symmetry axis and NCM which has now become 
the effective center of mass of the rotor(Fig.~\ref{fig3}). 

The problem can be described more clearly by considering the 
rotating hemisphere equivalently as a system of a circular 
disc which has the same mass, inertia and the rest
state center of mass as that of the rotating hemisphere. 
Obviously, rotating this disc would not cause
the shift of the center of mass. The unexpected effect of 
the anomalous center of mass shift
from the rotating hemispherical system(Fig.~\ref{fig1}) 
is the same as if the rest of the universe
exert uplifting force to the disc in addition to the usual 
outward normal force so that the
effective center of mass of the disc can be aligned with 
NCM while the center of the centripetal force in the disc 
is left at RCM
(Fig.~\ref{fig3}). The centripetal force is not capable
of balancing the vertical component of the force as it 
would with the radial component since
the vertical component of the force is not symmetrical 
inside the structure of the hemisphere.
If one assume that the center of the centripetal force 
must always be the same as the point
where the centrifugal force is centering around, there 
may not be the vertical component of
the force since the centrifugal and the centripetal 
force will be aligned in exactly the
opposite direction. But then there still remains the 
problem of explaining how an object
can jump from point A to point B and stop without any 
force being involved in the system.

The principle of inertial motion does not allow such type of 
movement for a massive object. 
In either explanations, it is obvious that there must 
exist a vertical component of force involved in the 
mechanics of this system. Either the RCM will try to get 
close to NCM or vice versa. If 
we choose the proposition that the tendency of the restoring 
force to return to the ground state makes the NCM to move 
toward RCM, the net result is that there remains non zero 
vertical component of force in the hemispherical system with 
respect to the rest of the universe. This linear force can easily be 
calculated by the triangular law (In Fig.~\ref{fig3}, 
$F_{centrifugal}$ pulls outward centering around NCM and RCM 
feels the internal stress of the vertical component of force 
toward NCM and the restoring force is opposite to that direction) 
for small shift of the center of mass, which is
given by \\ \\
\[F_{linear} = 
F_{centrifugal}(\frac{\delta r_{c}}{\sqrt{2/3} R})\] \\ \\
for $\omega R << c$, where $\sqrt{2/3} R$ is the 
effective radius where the total mass is imagined 
to be concentrated while giving the same inertia as that 
of the hemisphere for $\omega R << c$. This relation 
conforms the first rule by its explicit linear dependency 
on the anomalous center of mass shift.  The 
centrifugal force is given by \\ \\
\[F_{centrifugal} = 2\pi\sigma R^3 \int^{\pi}_{0} \frac{\omega^2
\sin^2{\theta}}{\sqrt{1 - \alpha \sin^2{\theta}}}
\,d\theta \cong \frac{\pi}{2}mR\omega^2\] \\ \\
for the hemispherical shell for $\alpha << 1$.

\ \\
\noindent\parbox{\textwidth}{\noindent\begin{fig}{fig3}
  \mbox{\setlength{\epsfxsize}{.80\textwidth} \epsfbox{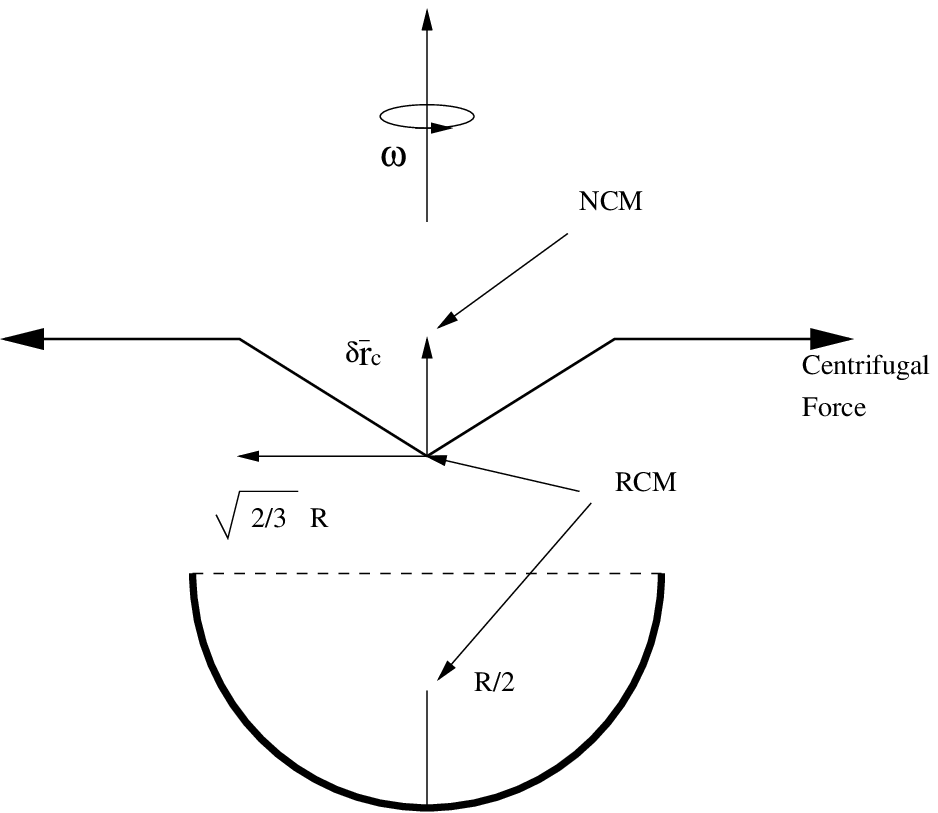} }\\
\ \\
  \figcap{.9\textwidth}{The shift $\delta r_{c}$ is 
enlarged for the illustration purpose. The shift in the center of mass 
causes the non zero component of the internal stress toward the 
positive z axis.}
\end{fig}
}\\

In this case, the linear force is given by \\ \\
\[F_{linear} = 
\frac{\pi m\omega^4 R^3}{48\sqrt{\frac{2}{3}} c^2}\] \\ \\
for  $R\omega << c$.
The force is proportional to the fourth power of the 
angular speed $\omega$ and to the third power 
of the radius of the hemisphere and the direction of 
the angular velocity does not affect 
the direction of the linear force. Calculation shows that 
the idealized perfectly rigid 
hemispherical shell of radius 1 m must rotate about 
550,000 rpm to generate acceleration g. 
The $\omega$ in this case is about 57,500/sec and 
$R\omega/c$ $1.92 \times 10^{-4}$. If the same system 
rotates at $R\omega/c$ equal to $1.92 \times 10^{-3}$ 
which is still in the non relativistic regime, the linear 
acceleration would become 10,000 g which is quite 
extraordinary. One may use an ultra centrifuge of 
1,000,000 rpm with a typical 20 cm diameter hemispherical 
rotor to test the increased gravity of 0.011g assuming 
the rotor is a hemispherical shell made of an 
idealized perfectly rigid material.
 
Since the angular momentum must be conserved for the 
axisymmetric rotating object, this continuous linear force 
is a mysterious one that could not have been deduced from
Newtonian mechanics. This local system gains energy as 
expected from its violation of Newton's first 
and third law of motion. After all, mechanics has been 
the basis of all of the thermodynamical energy conservation 
principles as demonstrated by the kinetic theory of gases. As
is well known, kinetic theory is based on the experimental 
verifications from the energetics of the 
confined ideal gases which are typically of spherical 
or longitudinal axially symmetric 
in shape. Therefore, this new principle has not been 
tested rigorously in this area. Even if 
this result is perplexing, the above conclusion seems 
unavoidable.

The origin of the centrifugal force and the inertia is not 
known well except that it may be from the action of the 
rest of the universe upon the local rotational motion of
an object according to the notion of Mach. Since the linear 
force which is caused by the anomalous center of mass shift 
is a vectorial part of the centrifugal force, the origin of
this force may also be attributed to that of Mach's. 
Newton himself once pondered that the centrifugal 
force may be the source of the gravitational force as shown 
in his experiment on a rotating bucket filled with water.

\bce\section*{\large\bf lV. Gravitational Dipole Moment}
\ece
\indent\indent

Now the question is ``What does this result have to do 
with the so far known results of 
general relativity ?''. General relativity allows the 
empty space solution even with the 
cosmological constant intact as shown by de Sitter. 
By his demonstration, it is proven that 
general relativity doesn't contain Mach's principle which requires the ever
present matter filled universe, contrary to Einstein's 
expectation. Because of this result, one
may suspect 
that the general relativistic equivalent of the Machian description of the
mechanics shown in 
this paper may not be found in general relativity.  

To investigate if this is indeed the case, consider the dipole term in the
linearized field 
equation of general relativity. The dipole term comes as the second term next to
the monopole 
field which is basically the source of Newtonian gravity in the multipole
expansion from the 
linearized weak field solution to general relativity. Since we are dealing with a
terrestrial 
system with sufficiently small mass and rotational velocity, the weak field
approximation 
should be enough for the investigation of the problem. 

In the near zone $(r << \lambda)$, but outside the source so that vacuum
Newtonian theory is nearly valid, 
the potential is \\ \\
\[\Phi = - \int_{all space} \frac{[T^{oo} + t^{oo} + T^{jj} + t^{jj}]_{ret}}{|x -
x'|}\,d^3 x' .\] \\ \\
For any nearly Newtonian, slow motion source, the condition \\ \\
\[|t^{oo} + T^{jj} + t^{jj}| << T^{oo}\] \\ \\
is satisfied. Therefore, one can write \\ \\
\[\Phi(x,t) = - \int \frac{[T^{oo}(x',t)]}{x - x'} \,d^3 x' .\] \\ \\	
The dipole term obtained from the expansion of the above potential is given by \\
\\
\[\Phi_{dipole} = - \frac{1}{r^3} (\int T^{oo} x_{j}^{'} \,d^3 x') x^{j}\] \\ \\
where the gravitational dipole moment \\ \\
\[\int T^{oo} x^{'}_{j} \,d^3 x'\] \\ \\
is equal to the total mass times the anomalous center of mass shift $( M\delta
\overline{r}_{c})$ calculated 
previously where the origin of the coordinate system is set at the rest state
center of 
mass of the hemisphere. 

The mass-energy density of the object depends on the total mass-energy of the
individual 
mass components comprising the rotating source. If one tries to be accurate in
describing 
the system even to the level of taking into account of the kinetic energy of each
individual 
mass components comprising the spatially extended object, one must include the
mass increase 
due to the special relativistic mass increase effect in the total mass-energy.
This is 
equivalent of taking the explicit account of the internal energy $U(\omega,r)$ in
addition to the 
rest mass-energy $m_{o}(r)$ and the gravitational potential energy $\Omega (r)$
since the second major 
term in the expansion of the relativistic mass is the kinetic energy
$\frac{1}{2}mv^2$. In this case, 
the internal kinetic energy arises from the collective rotational motion of the
mass 
components comprising the source, which means that the mass-energy density
$T^{oo}$ can not be a 
constant, to be precise, for an idealized perfectly rigid rotating source. 

Therefore, one has to make sure that the special relativistic mass increase effect
is left in 
the volume integral of the dipole term no matter how small the rotational
velocity of the 
source may be, which has not been the usual practice in the multipole expansion
of the potential 
for slowly rotating sources. The importance of this treatment becomes obvious
when we choose a 
slowly rotating idealized perfectly rigid hemisphere as a source, since by
neglecting this 
effect, we are in fact throwing out any possibility of the anomalous center of
mass shift 
from the beginning except for the trivial displacement that can be eliminated by
the simple 
spatial translation of the coordinate system.

According to the standard view, this dipole term can be made to vanish because it
has been falsely 
assumed that one can ``always'' align the origin of the coordinate system to the
center of mass 
of the object by translating the coordinate system. This assumption holds only
for objects which 
have axisymmetry and also point symmetry with respect to the center of mass while
the object is 
in rotational motion along the symmetry axis and also for non rotating systems.
However, as we 
have shown so far, the assumption doesn't work for the idealized perfectly rigid
rotating 
hemisphere which develops the shift of center of mass without external action in
the perpendicular 
direction to the rotational plane independent of the choice of the coordinate
system. This 
effective center of mass changes continuously depending on the angular speed of
the source and 
aligning the origin of the coordinate system to the effective center of mass
would not eliminate 
this kind of dual structure in the center of mass. In fact, one can not choose
the origin of the 
reference frame depending on the angular speed of the source, since, by doing so,
the coordinate 
system loses its meaning as a reference frame. It is seen now clearly how
consistently the 
rotating hemispherical type source becomes a peculiar and baffling mechanical
system for both 
Newtonian mechanics and general relativity as shown above. It is obvious that
this energy 
dependent anomalous center of mass shift must be identified as the true source of
the dipole 
term in the multipole expansion of the linearized field equation, not the trivial
displacement 
that can be eliminated by simple spatial translation of the coordinate system. 

The absence of direct physical evidence for such force in a terrestrial
environment may have 
contributed to the total negligence of the dipole term from the beginning when it
appeared in 
the Newtonian limit of general relativity. Most of 
the terrestrial physical problems can still be explained by Newtonian mechanics
except for few 
esoteric phenomena which are related to the dipole gravity in cosmological scale,
for example, 
the jets from the black hole accretion disk and the observed anomalous red shift
which will be 
discussed later, unless one created an artificial dipole system in a terrestrial
environment on 
purpose.

The multipole gravitational field may now be written (with G restored) \\ \\
\[\Phi = -\frac{GM}{r} + \frac{GM\delta r_{c}}{r^2} \cos{\theta} + O(\frac
{1}{r^3})\] \\ \\	
where $\theta$ is the angle between $\overline{r}$ and the anomalous center of
mass shift vector 
$\delta \overline{r}_{c}$ and M the mass 
of the source. The positive sign in the dipole 
term comes from the fact that the shift is toward 
the negative z axis when the hemisphere is placed 
like a dome in the xy plane.  Contrary to the quadrupole radiation 
proposed by Einstein\cite{B7}, the dipole field 
can assume the static field configuration under the presence of counteracting
external fields 
without the loss of energy. An object placed above the rotor in Fig.~\ref{fig1}.
would be repelled from 
the dipole and one under it attracted according to the inverse $r^3$ dipole force
law for 
$r >> \delta r_{c}$. 

Note that this dipole field and the direction of the polarity are consistent with
the linear 
force acting on the hemispherical rotor derived previously from the totally
different concept. 
These are unexpected coincidences corroborate the evidence for both the linear
force and the 
gravitational dipole moment. This dipole moment would not be accelerated in an
empty universe, 
although it may generate its own gravitational dipole field around it, which is
also consistent 
with the Mach's principle since there will be no centrifugal force in the empty
universe. 
Evidently, we have a Machian equivalent mechanics in general relativity which is
directly 
related to the presence of the gravitational dipole moment.

\bce\section*{\large\bf V. Conclusions}
\ece
\indent\indent

Since the theory predicts the existence of the dipole field in general relativity
which is much 
stronger than the quadrupole radiation from the binary stars\cite{B8}, one can
perform a controlled 
test by putting a massive object near the symmetry axis of a spinning hemisphere
and measuring 
the force it receives from the rotor as a function of r and $\omega$. 
Since the gravitational dipole moment has the force line which resembles exactly
that of the 
electric or magnetic dipole moment, the logical extension of this dipole
gravitational field 
created by the anomalous center of mass shift along the symmetry axis in the weak
field regime 
into the strong field regime would be the creation of a worm hole in the extreme
limit $R\omega > c$ 
which has been found in the Schwarzschild metric which connects two universes by
the two funnel 
type holes attached to each other by their ends. It represents a type of an one
way traversable 
worm hole whose creation can not be separated from passing through it, in
contrast with the one 
created by an infinitely long spinning cylinder proposed by van Stockum\cite{B9}
which does not have 
such an embedded mechanism. The system will travel from zero to near the speed of
light before 
it may create a worm hole assuming that the rotor is made of such an ideal
material and structure 
that it can withstand the extreme stress of the centrifugal force and also that
it satisfies the 
exotic material hypothesis\cite{B10}. Since the gravitational dipole moment has
the force line which comes 
out of the top of the hemisphere (Fig.~\ref{fig1}) and spread around to go
back normal into the bottom of 
the rotating hemisphere, where the force line going into a source is defined as
the attractive 
force, a beam of light passing from the bottom to the top will be
defocused following the force line in the extremely strong field regime, which
is, in fact, the 
statement that the gravitational dipole moment satisfies the exotic material
hypothesis. 

To view the dipole gravity further in relation to the electromagnetic
phenomena, note that 
the definition of the center of mass contains a term length times mass which is
identical in form 
to the definition of the electric dipole moment. Since a spinning hemisphere
seemingly at rest 
looking from a distance has in fact the center of mass different from that of an
identical object 
without spin, one may view the spinning hemisphere as having developed the ``dual
center of mass'', 
the source of the gravitational dipole moment. Restorable energy is required to
separate the 
opposite electric charges from the neutral state to produce the electric dipole
moment. 
Restorable energy is also required to produce the ``dual center of mass'' to
create the 
gravitational dipole moment. The fact that there doesn't exist negative mass
(negative energy 
density) in the universe has contributed to the notion that there is no
gravitational dipole 
moment. However, it is noted that
the repulsive 
pole of the gravitational dipole moment gives the same effect of defocusing a
beam of light 
passing through it as the negative mass (exotic matter) would. The physics for
this 
gravitational dipole moment happens to be the same as if there were negative mass
at the 
position NCM and the usual mass at RCM of equal absolute amount, only if one does
not attempt 
to derive the total monopole mass from this analogy. To avoid this confusion,
this negative 
mass may aptly be named as the ``negative image mass'' which doesn't really have
the mass except 
its effect. In this analogous system, it is defined that the negative mass repels
normal mass 
while the same kind of mass attracts each other in contrast to the case of
electrostatic charges. 

In retrospect, general relativity did predict the presence of the gravitational
dipole moment. 
There simply was no corresponding experimental data in the classical level to
recognize the 
dipole term as a physically meaningful source. The effect is barely observable at
the extremely 
fast rotational speed of 100,000 rpm for a 20 cm diameter hemispherical rotor,
which is indeed 
a very fast rotational motion according to our common present day experiences but
still far down 
in the non relativistic regime as far as the special relativistic criterion of
the instantaneous 
speed at the rim of the rotor is concerned. A sphere or a flat circular disc
would not produce 
the net directional force no matter how fast it rotates, according to the present
theory. 

Concerning this effect, the results of the experiment performed by Hayasaka and
\\
Takeuchi\cite{B11} are closely related to the present theory\cite{B12}. However,
there are three factors which 
indicate that the present theory and their experimental result may represent a
different 
physical effect. In the present theory: 1. The force is proportional to the
fourth power of 
the angular speed of the rotor from the start. 2. The force is independent of the
rotational 
mode of the rotor (either clockwise or counterclockwise). 
3. The force depends on the asymmetry of the rotor. 
On the other hand, in the Hayasaka-Takeuchi experiment: 1. The weight reduction
depends linearly 
on the angular speed of the rotor. 2. The weight reduction depends on the
rotational mode of 
the rotor (cw or ccw). 3. The weight reduction is not claimed to be dependent on
the asymmetry 
of the rotor (the detailed shape of the gyros and the configuration of the motor
is not provided 
in their paper except that the figure suggests the gyros are cylindrical type
solid objects). 
Therefore, it is unlikely that the present theory and the experiment represent
the same physical 
effect concerning the unknown force generated by the rotational motion of the
rotor. Especially 
the linear dependency of both the angular frequency and the effective radius of
the rotor on 
weight reduction in their experiment doesn't seem to fit the dimensional
requirement for the 
force. And any similar attempt to try such an experiment could not have been the
controlled 
one enough to satisfy both conditions; the shape of the rotor and the rotational
speed for 
the prescribed effect, without guidance from a theoretical prediction.

As an another example, the two opposite jet streams coming out of the black hole
accretion 
disk can also be explained by this mechanism by considering the spinning black
hole as a 
system of two dipoles attached face to face on the flat side of the hemispheres. 
Depending on the strength of the dipole moment, the amplification of the 
oscillations of the particles along the z axis may exceed the attractive 
force of the monopole field. The observed jets are considered to be the 
manifestation of the particle's trajectories experiencing this force 
although the detailed mechanism may need to be clarified. The 
two opposite jets also conform to the fact that the polarity of 
the dipole moment is independent of the direction of the rotation 
of each hemispheres as predicted.

It is also noted that this mechanism has the potential to explain the anomalous
red shift 
by considering the blue shifted galaxy as one chunk of the rotating point
asymmetric body 
the rotation axis of which is pointing toward our own galaxy. In this picture, it
is possible 
for galaxies to move in any predetermined direction depending on its asymmetry
and also on the 
rotational speed at the time of its birth apart from the Hubble expansion. 

\def\hebibliography#1{\begin{center}\subsection*{References}
\end{center}\list
  {[\arabic{enumi}]}{\settowidth\labelwidth{[#1]}
\leftmargin\labelwidth	  \advance\leftmargin\labelsep
    \usecounter{enumi}}
    \def\newblock{\hskip .11em plus .33em minus .07em}
    \sloppy\clubpenalty4000\widowpenalty4000
    \sfcode`\.=1000\relax}

\let\endhebibliography=\endlist

\begin{hebibliography}{100}

\bibitem{B1} S. W. Hawking, Phys. Rev. D {\bf 37}, 904 (1988)
\bibitem{B2} Michael S. Morris, Kip S. Thorne, and Ulvi Yurtsever, Phys. Rev.
Lett. {\bf 61}, 1446 (1988)
\bibitem{B3} C. W. Misner, K. S. Thorne and J. A. Wheeler, Gravitation (Freeman,
San Francisco, 1973)
\bibitem{B4} John L. Friedman and Michael S. Morris, Phys. Rev. Lett. {\bf 66},
401 (1991)
\bibitem{B5} Z. Thirring, Phys. {\bf 19}, 33 (1918); and 22, 29 (1921)
\bibitem{B6} Issac, Newton, Philosophiae Naturalis Principia Mathematica 
(Londini, 1967(1687))
\bibitem{B7} A. Einstein, Sitzungsber, Preuss. Akad. Wiss. Phys. Math. K1 {\bf
1916}, 688 and {\bf 1918}, 154.
\bibitem{B8} J. H. Taylor, L. A. Fowler, and P. M. McCulloch, Nature {\bf 277},
437 (1979); J. M. Weisberg 
and J. H. Taylor, Gen. Relativ. Gravit. {\bf 13}, 1(1981)
\bibitem{B9} W.J. van Stockum, Roy. Soc. Edinburgh, Proc. {\bf 57}, 135 (1937)
\bibitem{B10} Frank J. Tipler, Phys. Rev. Lett. {\bf 37}, 879 (1976)
\bibitem{B11} Hideo Hayasaka and Sakae Takeuchi, Phys. Rev. Lett. {\bf 63}, 2701
(1989)
\bibitem{B12} John Sangster,  private communication

\end{hebibliography}
\end{document}